\input epsf
\documentstyle[12pt]{article}

        {\newpage
        \setcounter{page}{1}}

\renewcommand{\title}[1]{%
        {\begin{center}
        \Large\bf #1
        \end{center}}
        \vskip .3in}

\renewcommand{\author}[1]{%
        {\begin{center}
        #1
        \end{center}}}

\renewcommand{\abstract}[1]{%
        \begin{center}%
        {\vspace{1em}\vspace{0pt}\bf Abstract}%
        \end{center}%
        \noindent #1}

\renewcommand{\date}[1]{%
        \begin{center}%
        #1%
        \end{center}}

        {\end{thebibliography}}

\makeatletter
        \@addtoreset{equation}{section}%
\makeatother

\newcommand{\eqn}[1]{\label{eq:#1}}

\newcommand{\refeq}[1]{(\ref{eq:#1})}
\newcommand{\eq}{eq.~\refeq}

\newcommand{\beq}{\begin{eqnarray}}
\newcommand{\eeq}{\end{eqnarray}}

\newcommand{\tr}{\mathop{\rm tr}}
\newcommand{\Tr}{\mathop{\rm Tr}}

\renewcommand{\Re}{\mathop{\rm Re}}

\newcommand{\twi}{\widetilde}
\newcommand{\mybar}[1]%
        {\kern 0.8pt\overline{\kern -0.8pt#1\kern -0.8pt}\kern 0.8pt}
\newcommand{\sla}[1]%
        {\hskip2pt\raise.15ex\hbox{$/$}\kern-.57em \hskip-2pt#1}
\newcommand{\roughly}[1]%
        {\mathrel{\raise.3ex\hbox{$#1$\kern-.75em\lower1ex\hbox{$\sim$}}}}

\newcommand{\drawsquare}[2]{\hbox{%
\rule{#2pt}{#1pt}\hskip-#2pt
\rule{#1pt}{#2pt}\hskip-#1pt
\rule[#1pt]{#1pt}{#2pt}}\rule[#1pt]{#2pt}{#2pt}\hskip-#2pt
\rule{#2pt}{#1pt}}

\newcommand{\Ybox}{\raisebox{-.5pt}{\drawsquare{6.5}{0.4}}}
\newcommand{\Ybbox}{\mybar{\raisebox{-.5pt}{\drawsquare{6.5}{0.4}}}}%

\newcommand{\jref}[4]{{\it #1} {\bf #2}, #3 (#4)}

\newcommand{\NPB}[3]{\jref{Nucl.\ Phys.}{B#1}{#2}{#3}}

\newcommand{\PLB}[3]{\jref{Phys.\ Lett.}{#1B}{#2}{#3}}

\newcommand{\PRD}[3]{\jref{Phys.\ Rev.}{D#1}{#2}{#3}}

\newcommand{\PRL}[3]{\jref{Phys.\ Rev.\ Lett.}{#1}{#2}{#3}}

\def\CA{{\cal A}}

\def\CL{{\cal L}}
\def\CM{{\cal M}}
\def\CN{{\cal N}}

\def\smsu#1{\rm\!\!S\!U\!(\!#1\!)\!\!}
\def\one{1\mskip-6mu1}

\begin{document}

\begin{titlepage}
\begin{center}
{\hbox to\hsize{hep-th/9805218 \hfill  BUHEP-98-14}}

\bigskip
\bigskip
\bigskip
\vskip.6in

{\Large \bf Duality of Non-Supersymmetric}\\

\bigskip

{\Large \bf Large N Gauge Theories} \\

\bigskip
\bigskip
\bigskip
\vskip.4in

{\bf Martin Schmaltz}\\

\vskip.2in

{\small \sl Department of Physics

Boston University

Boston, MA 02215, USA }

\smallskip

{\tt schmaltz@abel.bu.edu}

\vspace{1.5cm}
{\bf Abstract}\\
\end{center}

\bigskip

Starting from Seiberg's electric-magnetic duality for supersymmetric
QCD, we construct dual pairs of non-supersymmetric gauge theories. 
This is accomplished by first taking the large N limit of supersymmetric
QCD and it's dual partner and then performing a special
``orbifold projection'' recently introduced by Kachru and Silverstein.
We argue that in the large N limit the two projected theories
remain dual. The non-supersymmetric gauge theories which can
be studied in this fashion have non-supersymmetric field content,
chiral fermions and exactly massless scalar matter.

\bigskip

\end{titlepage}

\section{Introduction}

In recent years many examples of four dimensional supersymmetric
gauge theories have been found to exhibit non-Abelian
electric-magnetic duality \cite{seiberg,intrseib}.
The existence of a weakly coupled
dual description of the infrared physics of a strongly coupled
gauge theory allows one to obtain a wealth of information
on the low energy dynamics practically for free.

Clearly, it is very interesting to explore if electric-magnetic duality
also occurs for non-supersymmetric non-Abelian gauge theories.
Unfortunately only little progress in this direction has been made.
Several groups have introduced soft supersymmetry breaking to
supersymmetric dual pairs of theories in an attempt to deform
away from the supersymmetric limit in a controlled fashion.
However, intrinsic to this approach is that once the scale
of supersymmetry breaking is increased to be near the scale
of the strong dynamics noncalculable corrections become large
and make predictions impossible. Other attempts at guessing
non-supersymmetric dualities based on 't Hooft anomaly matching \cite{john}
or intuition from string theory \cite{nbranes} led to a few candidates of
non-supersymmetric duals. However in the absence of
decisive consistency checks these dualities remain suspect at best.

In this paper we propose a new approach to generating non-supersymmetric
duals. This approach employs a new technique which is motivated
from recent advances in string theory. Namely, Madacena \cite{malda}
conjectured that four dimensional $\CN=4$ superconformal $SU(N)$
gauge theory is dual to supergravity on five-dimensional anti-de Sitter
space in the limit of large $N$.
Kachru and Silverstein \cite{KS} extended the conjecture
to theories with less supersymmetry by orbifolding the supergravity
theory and determining the corresponding ``orbifold'' field theory.
Following up on this idea \cite{harvard} gave a general
recipe on how to generate ``orbifold projected'' daughter field
theories from the $\CN=4$ supersymmetric parent theory.
Using string perturbation techniques they further proved
the following statement to all orders in perturbation theory:

{\it the correlation functions of the ``orbifold'' daughter theories
are identical to corresponding correlation functions in the
parent theory in the limit of large $N$.}
This statement will be central to our derivation of
non-supersymmetric duality.
Finally, Bershadsky and Johansen \cite{BJ} showed how the
above statement can be proven in the context of large $N$ field
theory with no reference to string theory. They also
noted that the ``orbifold projection'' technique can
be applied more generally to large $N$ field theories with less than
$\CN=4$ supersymmetry.

The precise definition of orbifolding will be given in section
3. Roughly, the procedure is to identify a discrete global symmetry
of the parent field theory. One then also embeds this discrete symmetry
into the gauge group by using a special representation of the discrete
group. Now, ``orbifolding'' simply means
to eliminate all the fields of the parent theory which are not
invariant under the discrete symmetry. The daughter
theories interactions are inherited from the parent theory by
keeping all terms of the Lagrangian which only involve daughter fields.

In this paper we apply ``orbifolding'' to supersymmetric
QCD and its electric-magnetic dual to generate new non-supersymmetric
duals. The idea is simple: ``orbifolding'' takes a large $N$ parent
field theory and generates a daughter field theory whose
correlation functions are identical to the parent's. The daughters
are also large $N$ gauge theories but typically have less
or no supersymmetry. Applying
identical orbifolds to both the electric and magnetic theory
of Seibergs supersymmetric QCD generates two daughters.
Since the infrared properties of the parents were related
by duality, we conclude that the daughters must also be
related by duality. Note that we have ``derived''
non-supersymmetric duality from Seiberg's supersymmetric duality.

A possible pitfall which could invalidate the derivation is
that Seibergs duality might not capture large $N$ physics.
This is possible because the infrared limit in which
the duality is expected to hold might not commute with the large
$N$ limit. This would be the case if supersymmetric QCD includes
states with masses of order $\Lambda_{QCD}/N$ which are not included
in Seibergs dual, but which become massless in the limit of large $N$.
Such a situation does occur in QCD where the mass of the $\eta'$
goes to zero at large $N$ \cite{etaed}.
Such incompatibility of limits has
also been argued to occur near the monopole points of $\CN=2$ theories
\cite{ds}. In section 2 of this paper we study this concern by checking $\CN=1$
supersymmetric infrared results against large $N$ predictions
in two cases: the gluino condensate for supersymmetric Yang-Mills
theory \cite{shifvain}
and the strength of meson couplings in supersymmetric
QCD in the limit of large numbers of colors and flavors.
We find agreement with large $N$ expectations in both cases
and no evidence for new massless states spoiling the day.
Nevertheless, non-compatibility of the infrared and large $N$ limits
remains a worry and a more thorough investigation in future work
would be desirable.
 
The new duals which we will derive in this paper are very
far from being pairs of generic non-supersymmetric field theories. A
typical orbifold daughter and it's dual has a product gauge group of the
form $SU(d_1 N) \times SU(d_2 N)\times \ldots \times SU(d_n N)$
with gauge couplings given by
\beq
\eqn{coupl}
\left({g\over\sqrt{d_1}},{g\over\sqrt{d_2}},
\ldots,{g\over\sqrt{d_n}}\right)\ .
\eeq
Here $d_i$ are integers of order one
and $g$ is related to the running gauge coupling of the parent theory.
The daughters are often chiral and they typically have exactly massless
scalars in addition to the massless fermions. That the
scalars are kept massless to all orders in perturbation
theory is clearly a remnant of the supersymmetry in the
parent theory which would disappear if  -- for example -- we
allowed the ratios of couplings to differ from their finely tuned values
in \eq{coupl}.

It would be very exciting if the results of this paper
could be extended to finite $N$. There are two avenues to pursue:
one could try to orbifold parents with higher supersymmetry so
that the daughters still retain some supersymmetry, then
supersymmetric non-renormalization theorems can be used to argue
that the exact results continue to hold at finite $N$. An
example of such applications is the derivation of the Seiberg-Witten
curves for the coulomb branch of certain $\CN=1$ gauge theories
\cite{josh} by orbifolding $\CN=2$ theories \cite{erich}.
Another interesting avenue to pursue would be to use the
non-supersymmetric large $N$ duality as a starting point
to perturb around. Going to finite $N$ would introduce
corrections to the duality of order $1/N$ but one might still
find a qualitatively correct picture. At finite $N$ one
would expect the scalars to obtain a mass and possibly
also some chiral symmetry
breaking. It is conceivable that the symmetries and consistency with
the large $N$ limit turn out to be sufficiently constraining
to allow such a qualitative analysis.

Finally, it would be interesting to see if the orbifolds
considered in this paper can be related to brane configurations
in string theory or supergravity. One would expect to learn
more about both the field theories and string theory from such an
embedding.

The remainder of this paper is structured as follows. In section
2 we compare some of the exact supersymmetric results to large
$N$ expectations. In section 3 we introduce the method of
deriving a daughter field theory from a parent theory via
``orbifolding''. We also prove -- following \cite{BJ} --
that the correlators of the daughters are trivially related to
corresponding parent theory correlators. In section 4 we
apply the orbifold projection technique to Seiberg's electric-magnetic
dual pair to obtain an $SU(N)^3$ chiral
non-supersymmetric gauge theory and it's $SU(F-N)^3$
dual. We also perform a few consistency checks on this
new duality but we leave further exploration of the duality
and more extensive consistency checks to future work.

\section{Large N supersymmetric QCD}
In this section we compare expectations of the large-N expansion \cite{thoo}
with  exact infrared solutions to supersymmetric QCD due to
Seiberg and many others \cite{intrseib}.
In the examples we discuss we find agreement between the two techniques.
This result is not obvious because one might have worried that
the large-N theory contains states whose masses scale as
$\Lambda_{QCD}/N^{k}$ for some $k>0$
and become arbitrarily light in the large-N
limit. These states would be important to the infrared (IR) physics
of the large-N limit. However in the framework of the exact
supersymmetric results these states are excluded from
the effective IR Lagrangian because their masses are of
order $\Lambda_{QCD}$. Thus the two methods can give completely
different results if the large-N limit does not commute with
the IR limit taken by Seiberg and collaborators.
This problem appears to arise near the monopole points in
the Coulomb phase of $\CN=2$ supersymmetric theories as discussed in
\cite{ds}.
As we will show in this section, the limits do appear to commute
for supersymmetric QCD, allowing us to use both
techniques interchangeably. This result will be crucial to the
arguments of section 3 and 4.

\subsection{\it Gaugino condensate for supersymmetric glue}

In this section we compare large-N expectations for supersymmetric
QCD with no flavors with the exact results. The contents of this
subsection are not new and can be found in the literature
\cite{shifvain,nima,wittenmqcd,shif}.
We start with the Lagrangian in the customary normalization
 
\beq
\eqn{sqcd}
\CL = -{1 \over 4g_h^2} \int d^2\theta\  W_{\alpha} W^\alpha +{\rm h.c.}
    = -{1 \over 4 g^2}  (F_{\mu\nu})^2
    + {1 \over g^2} \bar\lambda i \sla{D} \lambda
    + {i\theta \over 32\pi^2}F_{\mu\nu}\tilde F^{\mu\nu} \ .
\eeq

Gauge fields $A_\mu$ and gaugino $\lambda$ transform in the adjoint
representation of the gauge group $SU(N)$, and a trace over
$SU(N)$ indices is implied.
The holomorphic gauge
coupling $g_h$ satisfies a simple renormalization group equation
which terminates at one-loop, and it is convenient to define the
renormalization group invariant holomorphic scale $\Lambda_h$ as

\beq
\eqn{holscale}
\Lambda_h^{3N} = M^{3N} e^{-8\pi^2/g_h^2(M)}\ .
\eeq

Here $g_h^2(M)$ is the holomorphic coupling evaluated at the
renormalization scale $M$. It is related to the conventional coupling $g$
of the Lagrangian with canonically normalized kinetic terms
via the Shifman-Vainshtein equation

\beq
\eqn{shifmanvainshtein}
\Re\left({8\pi^2 \over g_h^2}\right) = {8\pi^2 \over g^2} + N \ln g^2 \ .
\eeq

The large N limit of the theory is obtained by taking N to infinity
while keeping
\beq
\eqn{gscaling}
\bar{g}^2 = g^2 N
\eeq
fixed. In terms of the re-scaled coupling $\bar{g}$ the Lagrangian
takes the form

\beq
\eqn{sqcdscaled}
\CL = {N \over \bar{g}^2}\left[ -{1 \over 4} F_{\mu\nu}F^{\mu\nu}
    + \bar\lambda i \sla{D} \lambda\ \right] \ ,
\eeq

where a chiral rotation of the gaugino field has been performed to
set $\theta=0$. In this form the large N counting is performed
easily\footnote{
A lucid introduction to large N counting can be found in \cite{Sidney}.
Note that the usual large N arguments only apply to all orders in
perturbation theory, but it is expected that they continue
to hold non-perturbatively. The results of this section confirm
this expectation. }
with the usual well-known results. For the purpose of counting
N's gluinos can be treated identically to gluons.
The leading vacuum to vacuum amplitude is of order $N^2$ and
arises from planar diagrams with arbitrary numbers of gluon and gluino
loops. The expectation value of any operator which is
defined as a trace of elementary fields is of order N.
We therefore expect that the gluino condensate is of order N.
Let us check this prediction explicitly using the exact
result \cite{shifvain,seiberg}

\beq
\eqn{condensate}
< \tr \lambda\lambda > = e^{2\pi i k/N} \Lambda_h^3 \ .
\eeq

To continue we need to determine the $N$ scaling of $\Lambda_h$
which is obtained by exponentiating
\refeq{shifmanvainshtein}

\beq
\eqn{lambdascaling}
Re[\Lambda_h^3] = M^3\ {1 \over g^2}\ e^{-8\pi^2/g^2 N}
                           = M^3\ {N \over \bar{g}^2}\ e^{-8\pi^2/\bar{g}^2}
                           = N \Lambda^3 \ .
\eeq
The last equality defines $\Lambda$ which is both independent
of $N$ and renormalization group invariant. $\Lambda$ (and not
$\Lambda_h$) is the scale where the theory becomes strongly
coupled, and where one expects hadron masses. 
Thus we find that the gluino condensate scales like N as
predicted. Note that with canonically normalized
gluino fields the condensate scales as $N^2$.
Assuming confinement the large-N expansion further predicts
that glueballs and their superpartners are weakly interacting
with effective couplings of order $1/N$. Unfortunately we cannot
check this prediction because the massive hadron spectrum is
beyond the reach of our exact infrared techniques.

\subsection{\it Supersymmetric QCD with flavors}
 
In this section we consider supersymmetric QCD with $N$
colors and $F$ flavors in the large $N$ limit. There are
two qualitatively different large $N$ limits to consider
{\it i.} large $N$ with $F$ constant, in which case quark loops
are suppressed because there are less quarks
than gluons to run in loops, or {\it ii.} large
$N$ with $F/N$ constant in which case quark loops are just
as important as gluon loops.
The first alternative is the large $N$ limit which is most often
applied to QCD phenomenology. We will take $F$ large as well
because we are interested in comparing large $N$
expectations with the results of Seiberg's duality
which only exists for $F>N$.

The Lagrangian of supersymmetric QCD with flavors is
\beq
\eqn{sqcdflavors}
\CL = {N \over \bar{g}^2}\left[
-{1 \over 4}\int d^2\theta\  W_{\alpha} W^\alpha +{\rm h.c.}
+ \int d^4\theta\ (Q_i^\dagger e^V Q_i
+ \bar Q_i e^V \bar Q_i^\dagger)\ \right] \ ,
\eeq
where $Q$ and $\bar Q$ are the quark chiral superfields, $i=1,...,F$
is a flavor index, and we have scaled re-scaled all fields
such as to move the dependence on the gauge coupling $g^2=\bar g^2/N$
into the overall factor. It is convenient to perform the
$N$ counting in component form before integrating out the
auxiliary fields. Every propagator scales as $1/N$, every vertex
as $N$, loops of color lines as $N$, and loops of flavor lines
contribute $F\sim N$. It is then easy to see that the leading vacuum
to vacuum amplitudes are of order $N^2$, and are
given by planar diagrams with arbitrary numbers of
loops of adjoints and fundamentals.
If one assumes confinement one can estimate the interaction
strength of the confined degrees of freedom. For example for
properly normalized ``mesons'' $M=Q\bar Q$ one finds
\beq
\eqn{mesons}
<M_1...M_k>\sim N^{1-k/2}\ .
\eeq
Thus mesons interact weakly with effective coupling $N^{-1/2}$.
To compare this result with expectations from
Seiberg's duality we should keep in mind that we have assumed
confinement. Consequently we should only expect to find
agreement if the composite ``mesons'' are appropriate
low energy degrees of freedom.
One might also learn something interesting by studying
baryons which are more difficult to treat in the
large $N$ expansion; we leave this for future work.

The dual of supersymmetric QCD with $N$ colors and
$F$ flavors is an $SU(F-N)$ gauge theory with $F$ flavors
of dual quarks. There is also a fundamental ``meson'' which
is coupled to the dual quarks in the superpotential
$W=\mu^{-1} Mq\bar q$
where $\mu$ is the scale which appears in the matching
between the fundamental meson of the dual and the composite meson
of the electric theory $M=Q\bar Q$.
We now wish to take the large $N$ limit in the dual as
well. We take large $N$ and $F$ with
fixed ratio, but we also need to determine how to
scale the dual gauge and Yukawa couplings.

We can fix the scaling of the dual gauge coupling
by assuming that the magnetic theory has a sensible
large $N$ expansion. This determines that
$\twi g^2=\twi {\bar g}^2 N$. Demanding that loops with
internal mesons $M$ do not destroy the expansion
and that the meson Yukawa coupling does not become
irrelevant at large $N$ also fixes the large $N$
behavior of $\mu \sim N^{1/2}$.

Now we can compare with our prediction from the
eletric theory. There we found that the mesons
are weakly coupled with coupling of order $N^{-1/2}$
by assuming that the theory confines, that is, by assuming
that the mesons are proper infrared degreees of
freedom. From the dual description, we see that the mesons
are the correct infrared degrees of freedom for
$N < F <3/2 N$, and that their coupling is
$\mu^{-1} \sim N^{-1/2}$ as predicted.

\section{The ``orbifold'' projection}

In this section we discuss a projection technique which
allows one to generate pairs of field theories
which have identical large $N$
correlation functions. The theories which can be related
in this way generically have differing amounts of
supersymmetry, and often chiral theories are related to
non-chiral ones. One starts with a ``parent'' theory and
projects to a ``daughter'' theory by eliminating all fields
which are not invariant under a specifically chosen
discrete global symmetry of the parent Lagrangian.
Even though the technique is more general we will
limit ourselves to $SU(N)$ gauge theories with matter
in the adjoint and fundamental representations.
In this section we first discuss the technique of ``orbifolding''
in general, and then present three explicit examples
(and in section 4 we discuss the example of $SU(3N)$ orbifolded
by ${\bf Z}_3$ in detail).

The orbifolding technique was first introduced in the context
of Maldacena's conformal field theory/string theory duality
\cite{malda} by Kachru and Silverstein \cite{KS}
and was further developed and formalized in \cite{harvard}.
We will largely follow and expand on the work of
Bershadsky and Johansen \cite{BJ}
who re-derived the results of \cite{KS,harvard}
from the perturbative large $N$ expansion in field theory.
Before describing the projection technique we recall
a few basic group theory facts.

\subsection{\it A bit of group theory: regular representation and projectors}
A discrete finite
group $G=\{g_1,g_2,...,g_\Gamma\}$ with group multiplication
$\circ$ is associative, closed under multiplication,
has a unique identity element, and there exists a unique
inverse for each group element.
The {\it regular representation} of a group is given by
$\Gamma$ dimensional matrices $\gamma^a$
which are defined by $g_a \circ g_i = g_j (\gamma^a)_{ji}$.
Here the superscript $a=1,...,\Gamma$ labels group elements,
and there is no distinction between lower and upper indices
on the group elements.
The identity, $g_1$, is the $\Gamma \times \Gamma$ unit
matrix $\one_\Gamma$ in this representation.
Using the fact that
$g_a \circ g_b \ne g_b$ unless $a=1$
one immediately obtains that all group elements except the
identity element are traceless 
\beq
\eqn{traceless}
\Tr \gamma^a = \Gamma\ \delta^a_1\ .
\eeq
The regular representation is reducible: it can be shown that the
decomposition contains each irreducible representation
$R_l$ of the group with multiplicity equal to its
dimension $d_l={\rm dim}(R_l)$. Note that this property of
the regular representation implies the well-known
identity $\sum_l (d_l)^2=\Gamma$.
In a convenient basis the $\gamma^a$ are of the form
\beq
\eqn{gammablocks}
\gamma^a = \pmatrix{(r_1^a)&&&&\cr&(r_2^a)&&&\cr&&\ddots&
             \overbrace{\qquad\qquad\qquad}^{d_l \rm\;times}&\cr
             &&&\matrix{(r_l^a)&&\cr&\ddots&\cr
                        &&(r_l^a)\cr}          &\cr
             &&&&\ddots&\cr} \ .
\eeq
In the next section we will assign $\Gamma N$-dimensional
vectors (corresponding to the gauge indices of an $SU(\Gamma N)$
gauge group) to transform under an $N$-fold
copy of the regular representation.
These transformation matrices have the general form
\beq
\eqn{nfoldgamma}
\gamma_N^a = \bordermatrix{
  &\overbrace{\qquad\qquad\quad}^{d_1 N\rm\;times}&
             \overbrace{\qquad\qquad\quad}^{d_2 N\rm\;times}&&\cr
  &\pmatrix{(r_1^a)&&\cr&\ddots&\cr&&(r_1^a)\cr}&&&\cr
  &&\pmatrix{(r_2^a)&&\cr&\ddots&\cr&&(r_2^a)\cr}&&\cr
  &&&\ddots&\cr&&&&\ \ \cr} \ .
\eeq
The fundamental $Q$ and adjoint $A$ of $SU(\Gamma N)$ then
transform as
\beq
\eqn{adjointtrafo}
Q \rightarrow \gamma_N^a\ Q\ , \qquad
A \rightarrow \gamma_N^a\ A\ (\gamma_N^a)^\dagger \ .
\eeq
We will be interested in the components of the
matrix $A$ which are invariant under all such transformations.
They are easily
determined by applying Schur's Lemma which states that
a matrix which commutes with all elements of an irreducible
representation is a multiple of the unit matrix. We
find that the invariant components of $A$ are located in
blocks on the diagonal, transforming as an adjoint of
$SU(d_1 N)\times SU(d_2 N)\times \ldots \times SU(d_n N)
\subset SU(\Gamma N)$
\beq
\eqn{inva}
\pmatrix{A_1 \otimes \one_{d_1}&&&\cr
         &A_2 \otimes \one_{d_2}&&&\cr&&\ddots&\cr
         &&&A_n \otimes \one_{d_n}\cr} \ .
\eeq

We will also have use for a projector onto invariants of
the group. It is defined as
\beq
\eqn{projector}
P_R = {1\over \Gamma} \sum^\Gamma_{a=1} r^a\ , 
\eeq
where the $r^a$ are representation matrices of the (not
necessarily irreducible) representation $R$. Using
\beq
\eqn{gammap}
r^b\ P_R\ =\ {1\over \Gamma} \sum^\Gamma_{a=1} r^b\ r^a\ 
        =\ {1\over \Gamma} \sum^\Gamma_{c=1} r^c\ =\ P_R
\eeq
it is easy to show that $P^2=P$, and that $P=1$ in the trivial
representation, whereas $P=0$ in all other irreducible
representations. Thus when acting on a column vector
transforming in the representation $R$ the projector
extracts the invariant components.

Finally recall that one can form tensor product
representations $S \otimes T$ with representation 
matrices ${s^a \otimes t^a}$. The projector
onto invariants in such a tensor product
representation is
\beq
\eqn{tprojector}
P_{S\otimes T} = {1\over \Gamma} \sum^\Gamma_{a=1} s^a \otimes t^a\ . 
\eeq

\subsection{\it How to ``orbifold'' a field theory}

Using the regular representation and the projector 
defined in the last subsection we now define the
``orbifold'' projection which takes us from a large
$N$ parent theory to a daughter theory.
We consider a theory with gauge
group $SU(\Gamma N)$. We define the action of the discrete
group $G$ on all the fields of the theory by the following
procedure:
assign the gauge indices to $N$-fold copies of the regular
representation as in \eq{nfoldgamma}. A vector or an adjoint
then transform as in \eq{adjointtrafo}. Futhermore, we can
also embed the discrete group in the global symmetries
of the theory. If the size of a factor of the global symmetry
group grows with $N$ we assign the global group index to a
multiple of the regular representation as well. Global
groups under which the fields transform only in finite
dimensional representations
can be assigned to arbitrary representations of $G$.
The orbifolded theory is obtained by simply deleting
from the parent theory the fields which are not invariant under the action
of each element of $G$. The Lagrangian of the daughter theory is
obtained from the parent Lagrangian by keeping
the interactions which involve only invariant fields and
discarding all others.
In the next subsection we will prove that the correlators
of these daughter theories as defined here have the
very special property that their correlators are
identical to the correlators of their parents (after
a rescaling of the gauge coupling constants by a group theoretical
factor).
Before moving on, it is probably useful to demonstrate the
orbifolding procedure on three examples.

\noindent{\it Example i.}

First, consider $SU(\Gamma N)$ pure gauge theory.
The gauge indices are transformed by multiplication with
$\gamma^a_N$ as defined in \eq{nfoldgamma}.
Gluons in the adjoint of $SU(\Gamma N)$ transform as
$A_\mu \rightarrow \gamma_N^a\ A_\mu\ (\gamma_N^a)^\dagger$.
The invariant gluons in $A_\mu$ form an adjoint representation of
$SU(d_1 N)\times SU(d_2 N)\times\ldots\times SU(d_n N)\subset SU(\Gamma N)$.
Thus the daughter theory consists of $n$ decoupled $SU(d_l N)$
pure Yang-Mills theories. The interactions of the daughter theory
are obtained from the parent theory by taking the parent Lagrangian
and removing all terms which involve fields that were projected out.
In this case it is easy to see that the resulting Lagrangian
describes a product of decoupled Yang-Mills theories.
There is a small subtlety regarding the Yang-Mills couplings
of the $SU(d_i N)$ factors: they are not all equal to the
couplings of the parent theory. To see this recall that
the $i$'th invariant component of $A_\mu$ multiplies a $d_i$
dimensional unit matrix (\eq{inva}).
The traces over colors in the Lagrangian therefore include
also traces over these unit matrices $\one_{d_i}$ which each
yield a factor of $d_i$. Therefore, the projected theory has
overall factors of $d_i$ in front of the Lagrangians for
each of the Yang-Mills factors. By redefining couplings
and rescaling fields we can remove these factors. The end result
is that $SU(\Gamma N)$ with gauge coupling $g$ is
projected to $SU(d_1 N)\times SU(d_2 N)\times\ldots\times SU(d_n N)$
with gauge couplings
\beq
\eqn{couplings}
\left(g_1,g_2,\ldots,g_n\right)=\left(
{g\over\sqrt{d_1}},{g\over\sqrt{d_2}},\ldots,{g\over\sqrt{d_n}}\right)\ .
\eeq

Note that these couplings are not renormalization group invariant
but that ratios of couplings are invariant in the limit of
large $N$.
The statement that daughter and parent have identical correlation
functions in the large $N$ limit turns out to be trivial for
this example. The statement here is that correlators of $SU(d_i N)$
gauge theory with coupling ${g \over \sqrt{d_i}}$ are identical
to correlators of $SU(\Gamma N)$ with coupling ${g \over \sqrt{\Gamma}}$.
We knew that already since the large $N$ limit only
depends on the combination $g^2 N$ in either case.
However, this example can be viewed as a useful consistency check on
the arguments which we present in the next section.

\noindent{\it Example ii.}

Consider, as a second example, $\CN=1$ supersymmetric
$SU(\Gamma N)$ Yang-Mills theory. This theory has a global $U(1)_R$
symmetry which rotates the gluino field by a phase\footnote{In the
quantum theory the R symmetry
is broken by instantons to ${\bf Z}_{2 \Gamma\! N}$.
Strictly speaking, we therefore need to embed our global symmetry
into this subgroup. Similar comments apply to $U(1)_R$ and axial $U(1)$
in the next example.}. We can now embed the discrete group $G$
into the R symmetry such that the gluinos transform in a one dimensional
representation $R_l$ of $G$ in addition to their transformation
from the gauge indices. The gauge bosons do not carry R charge
and transform according to their gauge charge only
$A_\mu \rightarrow \gamma_N^a\ A_\mu\ (\gamma_N^a)^\dagger$.
Gluinos carry both gauge and R charge, they transform as
$\lambda \rightarrow r_l^a\ \gamma_N^a\ \lambda\ (\gamma_N^a)^\dagger$,
where $r_l^a$ is a transformation matrix in a one
dimensional representation, in other words, $r_l^a$ is a phase.
The invariant gluons are the same as in the first example. Therefore
the gauge group of the daughter is again
$SU(d_1 N)\times SU(d_2 N)\times\ldots\times SU(d_n N)$.
The gluinos are more interesting. Depending on which representation
$R_l$ we choose, different components of $\lambda$ survive
the projection. In the special case where $R_l$ is taken
to be the trivial representation the gluinos are projected
identically to the gluons, and we obtain $n$ decoupled $\CN=1$
supersymmetric $SU(d_i N)$ gauge theories.
In the more interesting case where $R_l$ is chosen to be
non-trivial we find a nonsupersymmetric daughter. Consider
for example the discrete group ${\bf Z}_\Gamma$ and pick the
representation
$R=\{r^a=e^{2\pi i a/\Gamma},\quad {\rm for}\ a=1\ldots\Gamma\}$.
With this choice the invariant gluinos are in $N\times N$ blocks
which are shifted to the right of the diagonal. The
resulting non-supersymmetric theory has chiral fermion field content
\beq
\eqn{exa2}
\begin{array}{ccccc}
SU(N)&SU(N)&SU(N)&\ldots&SU(N) \\[.1in] 
\Ybox &\Ybbox &1&\ldots&1  \\
1&\Ybox &\Ybbox &\ldots&1  \\
1&1&\Ybox &\ldots&1  \\
\vdots&\vdots&\vdots&\ddots&\vdots \\
\Ybbox &1&1&\ldots& \Ybox \\
\end{array}
\eeq
The interactions are determined by $SU(N)^\Gamma$ gauge invariance
with the gauge couplings of all the $SU(N)$ factors identified.
The overall normalization of the gauge couplings is identical to
the normalization in the original theory because the irreducible
representations of ${\bf Z}_\Gamma$ are all one dimensional. So the
$d_i$ which appear in \eq{couplings} are all equal to $1$.
A convenient way of representing the matter content of this
theory and similar orbifold theories is to use ``Moose''
\cite{Howard,nick} or ``Quiver'' diagrams \cite{quivers}.
The Moose diagram corresponding to \eq{exa2} is given in Fig 1.

\begin{figure}[t]
\centerline{\epsfxsize=2in \epsfbox{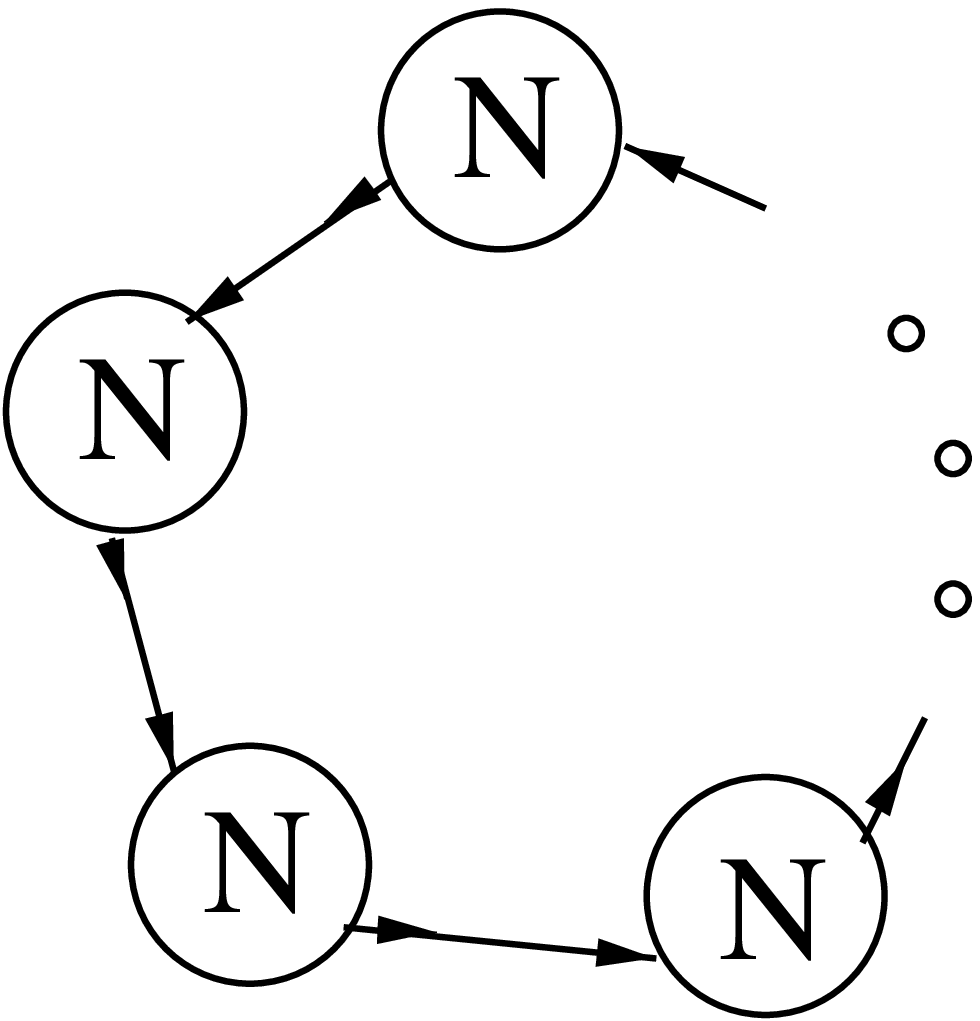}}
\vskip .2in
\noindent
Fig 1. {\it Diagram corresponding to an
$SU(N)\times SU(N)\times \ldots \times SU(N)$ gauge theory with
fermions in chiral representations. Each circle represents a
gauge group factor, and straight lines with arrows correspond to
fermion representations. Outgoing arrows indicate a fundamental
whereas ingoing arrows denote antifundamentals. For more details
on this graphical notation and it's application to duality
see \cite{nick}.}
\vskip .1in
\end{figure}

The result we have obtained for this theory is much less trivial.
We have related the chiral non-supersymmetric theory \eq{exa2}
to $N=1$ supersymmetric Yang-Mills, a theory about whose
vacuum structure we know a lot more.

\noindent{\it Example iii.}

Consider, as a third and final example,
$\CN=1$ supersymmetric
QCD with $\Gamma N$ colors and $\Gamma F$ flavors in the limit of
large $N$ and $F$  with $F/N$ fixed. This model requires a straightforward
generalization of the orbifold procedure described in \cite{BJ}.
In addition to using the regular representation to act on
gauge indices we also embed the discrete group into the
$SU(\Gamma F)_L \times SU(\Gamma F)_R$ global symmetry using an
$F$-fold copy of the regular representation. We define the
action of the $SU(\Gamma F)$'s such that all fields of supersymmetric
QCD transform with one upper and one lower index. Therefore
fundamentals of $SU(\Gamma N)$ are antifundamentals of
$SU(\Gamma F)$ and vice versa. Having only fields with one
upper and one lower index is not necessary but makes the
presentation more transparent.

The theory also has $U(1)_R$, Baryon number and axial $U(1)$ symmetries.
As in the previous example some of these symmetries
are anomalous so that we are really only dealing with
a discrete subgroup. We will nevertheless
continue to refer to these subgroups by
$U(1)$ whenever the context makes things clear.
We define a ``canonical'' $U(1)_R$ such that
gluinos have charge one and squarks are chargeless;
supersymmetry then determines the charge of quarks to be $-1$.
In order to perform the projection, we assign
axial $U(1)$ and Baryon number to the trivial representation of $G$,
and transform R charged fields in a one dimensional representation
as in the previous example.
We will use $Q, \bar Q$ and $\Psi, \bar\Psi$ for the scalar and
fermionic components of the quark superfields transforming in the
fundamental and antifundamental representations of $SU(\Gamma N)$
color.
The fields then transform as follows under $G$:
gluons and gluinos transform
as before, $A_\mu \rightarrow \gamma_N^a\ A_\mu\ (\gamma_N^a)^\dagger$,
$\lambda \rightarrow r_l^a\ \gamma_N^a\ \lambda\ (\gamma_N^a)^\dagger$.
Squarks transform as $Q \rightarrow \gamma_N^a\ Q\ (\gamma_F^a)^\dagger$
and $\bar Q \rightarrow (\gamma_N^a)^\dagger\ \bar Q\ \gamma_F^a$.
And finally quarks transform as
$\Psi \rightarrow (r_l^a)^\dagger\ \gamma_N^a\ \Psi\ (\gamma_F^a)^\dagger$ and
$\bar\Psi \rightarrow (r_l^a)^\dagger\ 
(\gamma_N^a)^\dagger\ \bar\Psi\ \gamma_F^a$.

Choosing as in the second example above the discrete group
$G={\bf Z}_\Gamma$ and representation
$R=\{e^{2\pi i k/\gamma},\quad {\rm for}\ k=1\ldots\Gamma\}$ for the
embedding into $U(1)_R$ we get the matter content depicted in Fig 2.
%
\begin{figure}[t]
\centerline{\epsfxsize=3.5in \epsfbox{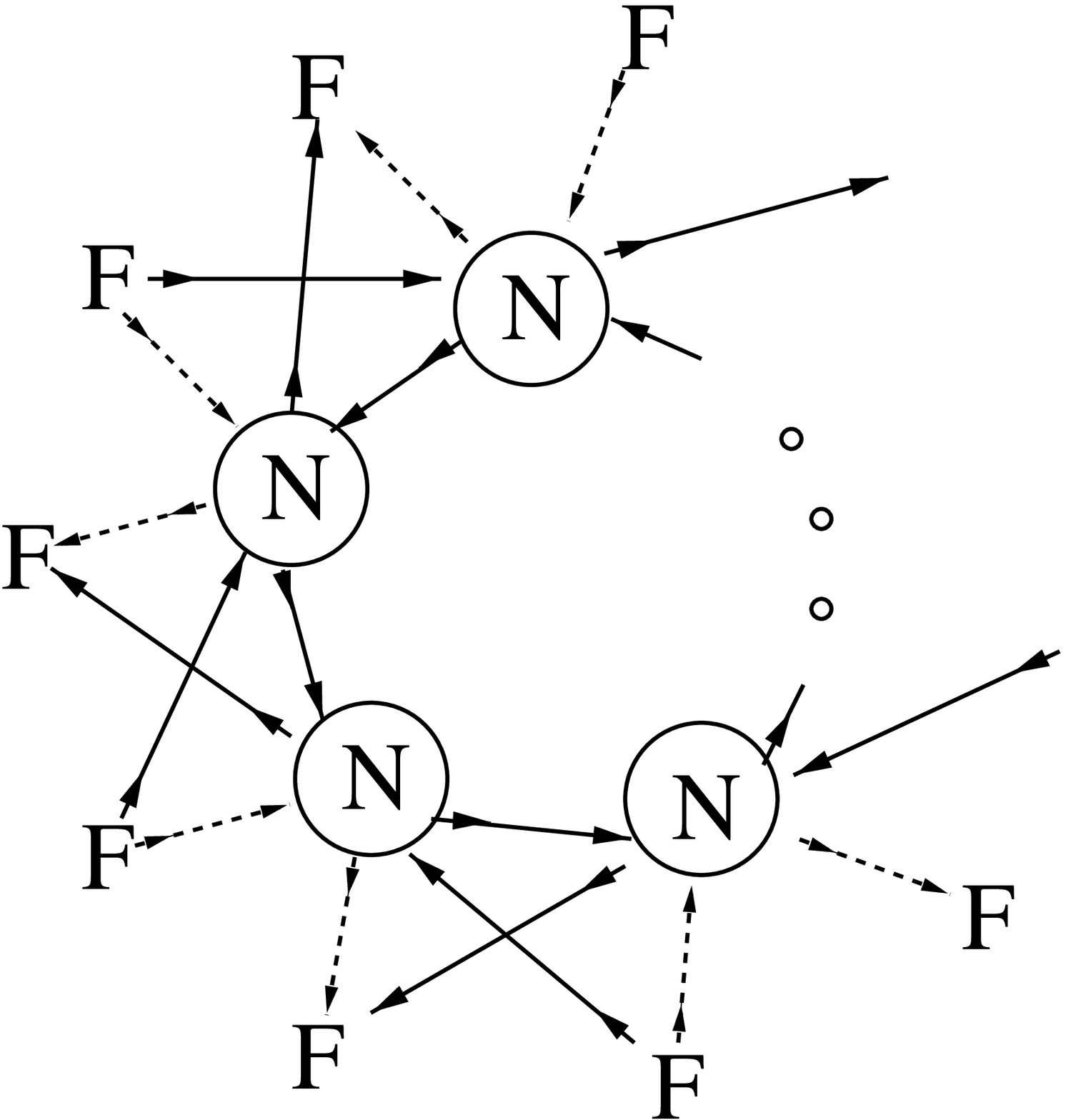}}
\vskip .2in
\noindent
Fig 2. {\it Diagram corresponding to an $SU(N)^k$ gauge
theory with $SU(F)^k\times SU(F)^k$ flavor group. Each
solid line with arrows indicates a fermion representation
while a dashed line indicates a scalar. The arrows indicate the
representation under the gauge or flavor group, outgoing arrows
stand for fundamentals and ingoing arrows for antifundamentals.}
\vskip .1in
\end{figure}
%
The couplings of the projected theory are equal to each other
because all of the $d_i$ in \eq{couplings} are equal to one.
The special case of $SU(3N) \rightarrow SU(N)\times SU(N)
\times SU(N)$ is discussed in detail in section 4.

\subsection{Large $N$ correlators of parent and daughter theories}

In this section we apply and extend the arguments of Bershadsky
and Johansen \cite{BJ} who proved that all
correlators of the daughter theories
are given by corresponding correlators of the parent theory
after a trivial rescaling of the couplings. The practical
application of this result is that one can calculate
a correlator of the (possibly non-supersymmetric) daughter
theory by instead doing the calculation in the full
supersymmetric parent theory and rescale the coupling constants in the end
\beq
\eqn{parentdaugh}
\CM_{daughter}(g^2_i={g^2\over d_i})=\CM_{parent}({g^2\over\Gamma})\ .
\eeq

Here $\CM_{daughter}$ is a correlator of the daughter theory
with some number of external fields. It depends on
the various coupling constants $g_i$ of the daughter theory
which have specail values as determined by the orbifold
projection procedure (see section 3.2, example 1).

We will first show that correlators between the daughter and
parent are related at {\it one loop} by evaluating
the color and flavor traces for a typical
one loop diagram in both theories explicitly. Then we extend the proof
to all orders by simply iterating the one-loop result.
For most of this section we will follow the proof in \cite{BJ} which
applies only to theories with fields in adjoint representations.
But as already predicted in \cite{BJ} the proof can easily be generalized,
which is what we present here for the case of bi-fundamental
matter fields, the case relevant for ``orbifolding''
supersymmetric QCD.

In the proof we will use 't Hooft's double-line
notation \cite{thoo}. In this notation the flow of color and flavor
indices through a Feynman diagram is made explicit.
A propagator for a field in the fundamental is represented
by a single line $\rightarrow\mskip-12mu-$
with an arrow indicating the direction of
the index flow. For simplicity we will limit ourselves
to theories with adjoints and bifundamentals. We
choose the representations such that all fields
carry one fundamental and one antifundamental index,
so that they are always represented by two lines with oppositely
oriented arrows 
$\>_{\rightarrow\!\!-}^{-\mskip-6mu\leftarrow}$.
This orientation
of the arrows is important for the arguments below.
\footnote{More general applications
are possible as well. For example, theories with symmetric
and/or antisymmetric tensors might be interesting to study.
Some care is required in extending the following arguments to
theories with such tensors as their index flow corresponds
to two lines with the same orientation.}
Furthermore, to simplify the $N$ counting we only consider
interaction vertices with color and flavor indices contracted
in a single trace. Such a vertex cannot be written as a product of
two separately color and flavor invariant operators.

Before presenting the chain of arguments which proves the
claim made above we list two general properties of large
$N$ diagrams which are used in the proof.
{\it i.} 
in the large $N$ limit the perturbation series is dominated
by planar diagrams with arbitrary numbers of loops.\footnote{
We will not discuss gauge choices and associated ghosts.
The large $N$ arguments as well as the orbifold projection
are easily extended with no fundamental changes to diagrams
with ghosts.}
Loops of adjoint or bifundamental fields contribute
equal $N$ factors as we are considering $F\sim N$.
{\it ii.}
all external (double-)lines should be attached to
a single index loop. This is true because the insertion
of external lines into a loop breaks the index flow in the loop
which costs a factor of $N$. Putting external
fields on more than one loop would cost additional powers of $N$.
Thus at leading order in $1/N$ all external fields
attach to a single quark loop which we can choose to be
the boundary of the diagram without loss of generality.

%
\begin{figure}[t]
\centerline{\epsfxsize=3.5in \epsfbox{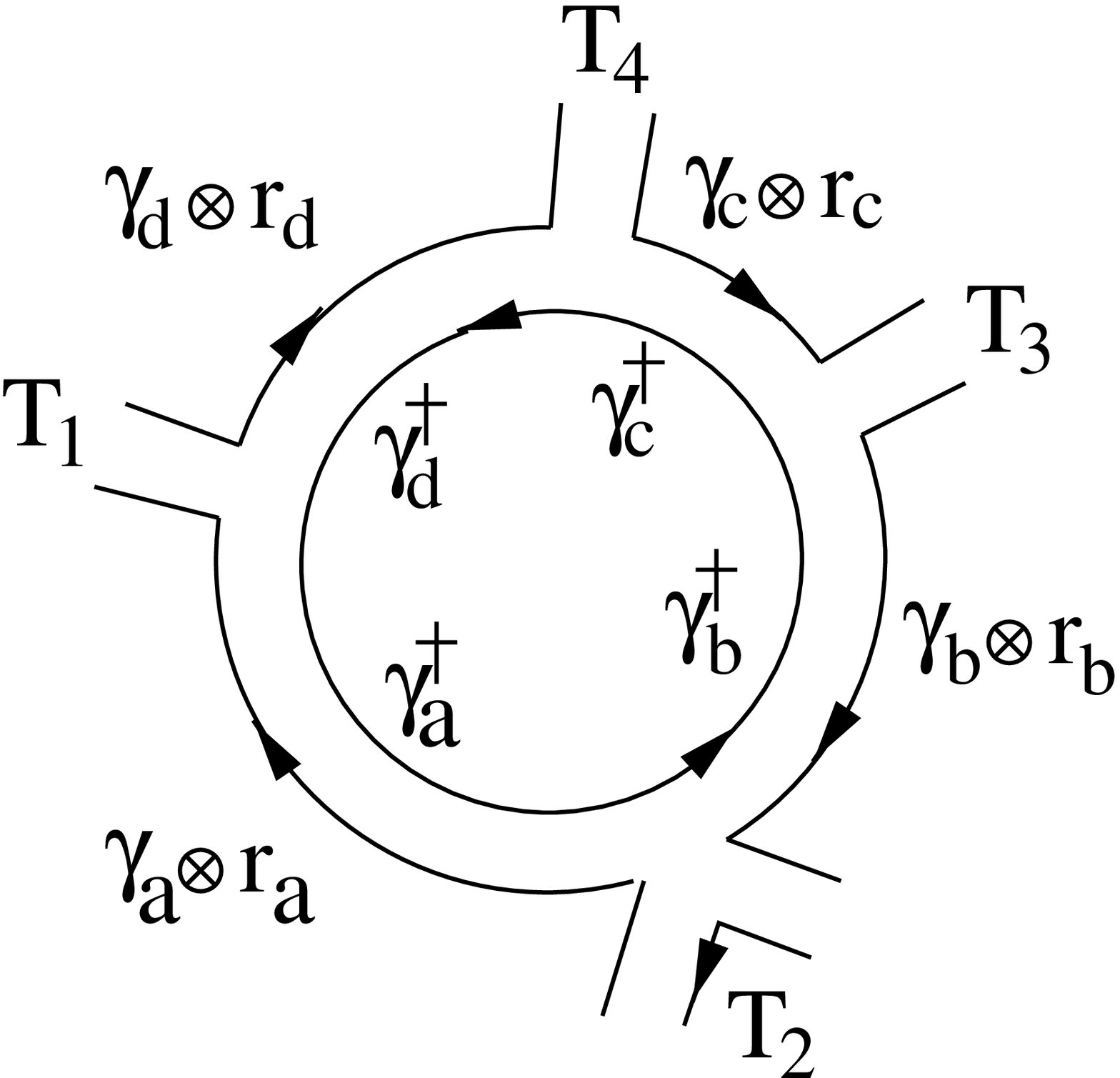}}
\vskip .2in
\noindent
Fig 3. {\it A typical one-loop diagram with projectors.}
\vskip .1in
\end{figure}
%
We now prove that \eq{parentdaugh} holds at one loop
by showing that it holds for simply connected one loop
diagrams contributing to $\CM$.
A typical planar one-loop diagram in the daughter theory with five
external fields is shown in figure 3. All the propagators in the diagram
correspond to fields of the daughter theory but we can rewrite the
diagram entirely as a diagram involving fields of the parent
theory with appropriate projectors inserted. 
The external lines are restricted to daughter theory fields but we
can rewrite them as parent theory fields and insert
a projector as in \eq{projector}
$P={1\over\Gamma}\sum\gamma_a \otimes r_a\otimes \gamma_a^\dagger$
which projects onto the daughter fields.
Here $\gamma_a$ and $\gamma_a^\dagger$ act on the indices
represented by the double lines (color and flavor)
whereas the $r_a$ act on the remaining indices which
transform nontrivially under $G$.
The internal propagators
are also for fields of the projected theory but again we can
instead use propagators of the parent theory and insert
a projector $P$ with every propagator.
The interaction vertices $T_1,...,T_4$ are all invariant
under $G$, and since all the lines coming into each vertex
are projected, the interaction vertices can also be
replaced by vertices from the parent theory.
We have now written the diagram entirely in terms
of parent theory propagators and vertices by
adding projectors for each internal propagator and external
line. Note that be rewriting the diagrams in this way we
relate the daughter theory with couplings $g_i=g/\sqrt{d_i}$
to the parent theory with coupling $g$. (See the discussion
in example of 1 of the last section.)

The projectors are represented in figure 3 by
the $\gamma^\dagger$ and
$\gamma \otimes r$. The $\gamma^\dagger$
act on the innermost index line whereas the $\gamma$'s act on the
exterior one. The $r$'s act on indices of $T$ which are not
displayed in the double line notation. For example, in the
$\CN=4$ supersymmetric theories considered in \cite{KS,harvard,BJ}
the $r$'s act on $SU(4)_R$ indices.
Finally the indices
$a,b,c,d$ are all summed from $1$ to $\Gamma$ and the diagram
has an overall factor of ${1\over \Gamma^4}$ from the projectors
on internal propagators\footnote{
Note that we have been careless by not designating which
index lines are flavor or color lines. This will be reflected in the
equations to come by an ambiguity over which representations
$\gamma$ and $r$ matrices are in. For example, the $\gamma$ could be
either in an $N$ or $F$ fold copy of the regular representation,
depending on whether they are inserted on a color or flavor index loop.
The difference is that closed flavor loops
give factors of $\Gamma F$ whereas color loops give $\Gamma\!N$.
We will ultimately only be interested in the ratio of amplitudes
in the daughter theory to amplitudes in the parent theory. In
the ratio the difference between color and flavor lines drops out.}.

It will be the goal of the next few paragraphs to show that this
daughter theory diagram is equal to ${1\over \Gamma}$ times
the corresponding parent theory diagram (which is obtained by leaving
out all the projectors).

First we remark that the diagram factors
into an overall numerical ``group theory factor''which we call
$\CA$, and the complicated rest of the diagram with all it's
spin and momentum dependence. We define the ``group theory diagram''
to contain the traces over color, flavor, and all other internal
indices as well as the coupling constants. Thus we can
represent the Feynman diagram as a product of two
diagrams: one ``group theory diagram'' which contains
the internal index structure and coupling constants for
which we use 't Hooft's double line notation,
and a ``stripped diagram'' which has been stripped
of all internal group structure but whose propagators carry
the momentum and spin dependence of the original Feynman diagram.
The difference between the daughter and parent
theory diagrams is entirely in the internal index structure.
We are interested in the difference between daughter
and parent diagrams for which the
``stripped diagram'' is irrelevant. Thus from now on we will
only be concerned with the ``group theory diagram''
and forget all momentum and spin dependence. It should be clear
that the factorization into a ``group theory diagram'' and
a ``stripped diagram'' is general and continues to hold for
all higher loop diagrams as well. We will use this fact
when we generalize to all orders.

We can now explicitly write the amplitude corresponding
to the ``group theory diagram'' of figure 3
\begin{eqnarray}
\eqn{amplitude}
\CA &=&g^5{1\over\Gamma^4}\sum_{a,b,c,d=1}^\Gamma
\Tr[\gamma_{a}^\dagger \gamma_{b}^\dagger
\gamma_{c}^\dagger \gamma_{d}^\dagger]\\ 
&&\Tr[T_1(\gamma_{d}\otimes r_{d})T_4(\gamma_{c}\otimes r_{c})
T_3(\gamma_{b}\otimes r_{b})T_2(\gamma_{a}\otimes r_{a})]\nonumber
\end{eqnarray}
The traces simply follow from following the index lines
in the direction of the arrows and contracting all
matrices which we encounter along the way in a big trace.
This is simple for the inner loop but for the outer
loop we must also include $T$ tensors which come from
the interaction vertices. The $T$'s carry a number of indices
for each double line entering the vertices; for each
double line there are two color or flavor indices and one
index that the $r$'s act on.
The $g^5$ prefactor comes from the coupling constants
at the interaction vertices.
In general there can be different coupling constants
for different types of vertices. We generically denote them
all by $g$. Again the difference will not matter in the end.
What is important is that we define the couplings such that
an $n$ field vertex has a coupling $g^{n-2}$. This prescription
is important for the rescaling of couplings which needs to be
done in the end. Note that this assignment is automatic for
gauge interactions, is consistent with supersymmetry,
and is required for the existence of a non-trivial large $N$
limit.

To evaluate amplitude \eq{amplitude} recall that the
$\gamma$'s are in multiples of
the regular representation whose generators are traceless except
for the identity generator.
Therefore the diagram vanishes unless 
\beq
\eqn{one}
\gamma_{a}^\dagger \gamma_{b}^\dagger
\gamma_{c}^\dagger \gamma_{d}^\dagger=\one_{\Gamma\!N}\ .
\eeq
We can use this fact to kill one of the sums, say over $d$
and set $\gamma_{d}=\gamma_{a}^\dagger
\gamma_{b}^\dagger \gamma_{c}^\dagger$.
Since the $\gamma$'s form a faithful representation of $G$,
\eq{one} also holds for group elements
$g_{a}^{-1} g_{b}^{-1} g_{c}^{-1} g_{d}^{-1}$=1.
Thus we also have
$ r_{d}=r_{a}^\dagger r_{b}^\dagger r_{c}^\dagger$.
This leaves us with

\begin{eqnarray}
\eqn{amplitude2}
\CA&=&g^5{1\over\Gamma^4}\sum_{a,b,c=1}^\Gamma \Tr[\one_{\Gamma\!N}]\\
&&\Tr[T_1(\gamma_{a}^\dagger\otimes r_{a}^\dagger)
(\gamma_{b}^\dagger\otimes r_{b}^\dagger)
(\gamma_{c}^\dagger\otimes r_{c}^\dagger)T_4(\gamma_{c}\otimes r_{c})
T_3(\gamma_{b}\otimes r_{b})T_2(\gamma_{a}\otimes r_{a})]\nonumber
\end{eqnarray}
In the next paragraph we prove that the $(\gamma\otimes r)$ can be commuted
through the $T$'s; we use this fact here to simplify \eq{amplitude2}
by moving all the $T$'s to the right of the $(\gamma\otimes r)$'s.
Then we can annihilate
all $\gamma\otimes r$ with $\gamma^\dagger\otimes r^\dagger$ and obtain
\beq
\eqn{amplitude3}
\CA=g^5{1\over\Gamma^4}\sum_{a,b,c=1}^\Gamma \Tr[\one_{\Gamma\!N}]\ 
\Tr[T_4T_3T_2T_1]\ .
\eeq
The sums have all become trivial and give factors of $\Gamma$
and we finally have
\beq
\eqn{amplitude4}
\CA=g^5 N  \Tr[T_4T_3T_2T_1]\ .
\eeq
This should be compared with the amplitude of the parent
theory which is almost identical: it has the same trace
$\Tr[T_4T_3T_2T_1]$ and coupling constants $g_{parent}^5$
but it has a factor of $\Gamma\!N$ from the closed loop.
The two amplitudes differ only by a factor of $\Gamma$
which we can absorb into the coupling constant $g^2$. 
After the rescaling
the loop counting parameter $\bar g^2$ is the same in
both theories, $d_i N \times {g^2\over d_i}$ in the daughter
and $\Gamma N \times {g^2\over\Gamma}$ in the parent.
Thus we have shown that \eq{parentdaugh} holds at one loop.

Before we get to the generalization to all loops we
need to justify that we were allowed to commute $T$'s
with $\gamma\otimes r$'s. To do this we first
put back explicit indices on the vertices $T^{LRIJK\ldots}$.
The $L$ and $R$ indices stand collectively for all the indices of the
internal propagators of the loop which enter into the vertex
from the left or right. The $IJK\ldots$ stand for the
external lines, one index for each external line.
Now recall that the interaction vertices
of the parent theory are $G$ invariant. Thus transforming
all indices of $T^{LRIJK\ldots}$ with an element
$g_a\in\Gamma$ leaves it invariant.
Doing such a transformation introduces
a $\gamma_a \otimes r_a\otimes \gamma_a^\dagger$ on each
line emerging from $T$. On the external lines these
matrices can be absorbed into the projectors $P$ using
\eq{gammap}. We then have
\beq
T^{LRIJK\ldots}=\left(\gamma_a^\dagger \otimes r_a^{(L)} \otimes
\gamma_a\right)_L\ T^{LRIJK\ldots}\ 
\left(\gamma_a \otimes (r^\dagger_a)^{(R)}\otimes \gamma_a^\dagger\right)_R
\eeq
where it is implied that the external lines $LRIJK\ldots$ are
all projected, and the $\dagger$'s on the $r_a$ are chosen to
conform with the conventions used above. It is obvious from
the double line notation (see figure 3) that one of the $\gamma$'s is directly
contracted with the $\gamma^\dagger$ and we can annihilate them.
So we finally find
\beq
\left(\gamma_a^\dagger \otimes (r^\dagger_a)^{(L)}\right)_L\ T^{LRIJK\ldots}=
T^{LRIJK\ldots}\left( (r^\dagger_a)^{(R)}\otimes \gamma_a^\dagger\right)_R
\eeq
which shows that we can commute the $\gamma_a \otimes r_a$
past the $T$'s as claimed. Note that in the process the
matrix $\gamma_a\otimes r_a$ changes from the representation appropriate to
act on the propagator to the left of $T$ in the loop to the one
appropriate for the propagator on the right.

%
\begin{figure}[t]
\centerline{\epsfxsize=3.5in \epsfbox{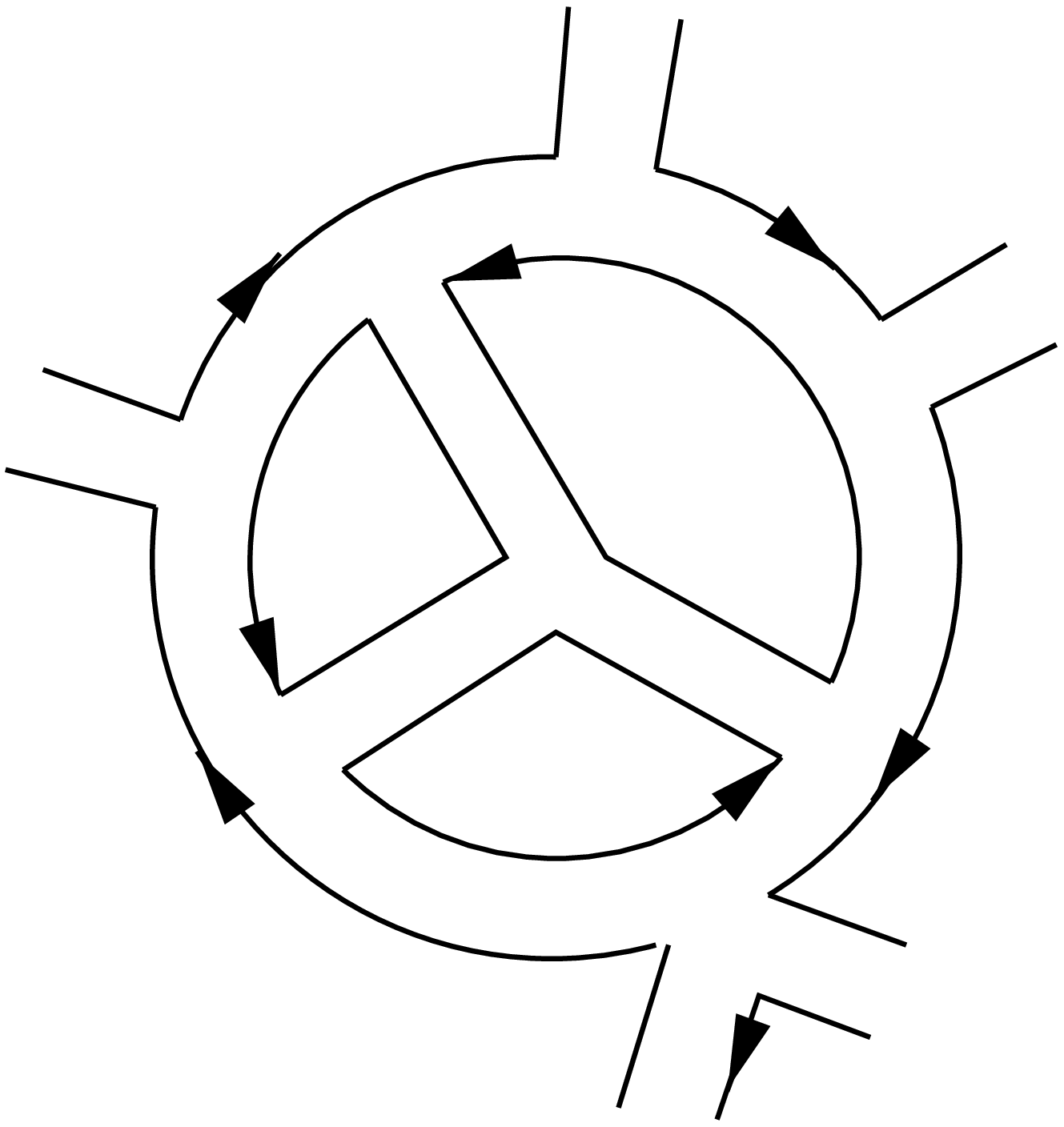}}
\vskip .2in
\noindent
Fig 4. {\it A typical planar three-loop diagram.}
\vskip .1in
\end{figure}
%
The generalization of the derivation of \eq{parentdaugh} to all
orders is very simple. The leading large $N$ diagrams
are planar with arbitrary numbers of color and flavor loops.
Again, a general planar diagram can
be split into a numerical group theory factor which we
represent by a double line diagram times a factor containing
all the spin and momentum information. The second factor
can be ignored because it drops out in the ratio of daughter
to parent. To compute the ``group theory diagram''
(see figure 4 for a typical three-loop example)
in the daughter theory we proceed exactly as before, we
rewrite the diagram entirely in terms of parent theory
propagators and vertices with projectors. Then we choose
one of the internal index loops and shrink it to
an effective vertex by calculating the trace for this
loop. Note that this calculation is identical to the calculation
we just did for the one-loop derivation. We obtain a new diagram
with one fewer loop and a new effective vertex with an
effective coupling. As before we find that the effective
vertex is the same as what we would have obtained if
we had calculated in the parent theory with a coupling constant
rescaled by ${1\over \sqrt{\Gamma}}$.
We now iterate this procedure by shrinking loop
after loop. Each time we find that we get the same answer as
in the parent theory with the rescaled coupling. This
completes the derivation of \eq{parentdaugh} to all orders in
perturbation theory.

One might worry that the equality of correlation functions
of daughter and parent theories \eq{parentdaugh} could
be spoiled by non-perturbative effects which would then
invalidate the dualities we derive. This
worry about non-perturbative effects is not special to
our situation, and we take the successes of large $N$ expansions
in general as evidence that large $N$ perturbative arguments
also apply non-perturbatively. The agreement
of the non-perturbative gluino condensate calculation with large
$N$ expectations (see section 2) corroborates this argument, and we will assume
that \eq{parentdaugh} continues to hold non-perturbatively.

\section{Example: chiral $SU(N)\times SU(N)\times SU(N)$}

In this section we construct our explicit example of a
non-supersymmetric dual pair. We begin by reviewing Seiberg's
electric-magnetic duality. Then we apply the orbifold projection
to obtain the field content and interactions of the
non-supersymmetric dual pair. We conclude with a few comments
and perform consistency checks on the duality.

\subsection{\it Parent theory duality}

As parent theory choose $N=1$ supersymmetric QCD with $3N$ colors
and $3F$ flavors. The non-anomalous global symmetries of this
``electric'' theory are summarized in the following table

\beq
\begin{array}{c|c|cc|rr}
        & SU(3N) &SU(3F)&SU(3F)& B & R \\[.1in]\hline
&&&&&\\[-.1in]
Q   & \Ybox    & \Ybbox & 1      &{1\over N} &{F-N \over F}\\
\bar Q& \Ybbox   & 1      & \Ybox  &-{1\over N}&{F-N \over F}\\
\end{array}
\eqn{electricqcd}
\eeq
\medskip

where $Q$ and $\bar Q$ denote the quark superfields. We will
always denote the scalar component of a chiral superfield
with the same letter as the superfield. Thus for the squarks
we will also use $Q$ and $\bar Q$, and we will use
$\Psi$ and $\bar\Psi$ for the fermionic components
of the quark superfields.
The vector superfield which contains the gluons $A$
also contains an adjoint of fermions, the gluinos $\lambda$.

The two nonanomalous $U(1)$
symmetries can be determined by assigning charges to the
three fermion fields $\lambda, \Psi, \bar\Psi$ subject to the
$SU(3N)^2 U(1)$ anomaly cancellation constraint. The charges of the
scalars are then determined by supersymmetry. All the
global symmetries except the $R$ symmetry commute with
supersymmetry, implying that boson and fermion components
of the superfields have identical charges. The $R$ charges
of bosons and fermions differ because the superspace coordinate
$\theta$ transforms with charge one.

Supersymmetric QCD has an IR-dual with $3\twi N \equiv 3(F-N)$
dual colors and $3F$ flavors. The fields of this ``magnetic'' theory
transform as

\beq
\begin{array}{c|c|cc|rr}
        & SU(3\twi N) &SU(3F)&SU(3F)& B & R \\[.1in]\hline
&&&&&\\[-.1in]
q  & \Ybbox    & \Ybox & 1      &{1\over \twi N} &{N \over F}\\
\bar q& \Ybox   & 1      & \Ybbox  &-{1\over \twi N}&{N \over F}\\
m       & 1        & \Ybbox & \Ybox  & 0         &2{F-N \over F} \\
\end{array}
\eqn{magneticqcd}
\eeq
\medskip

Here $q, \bar q$ are the dual quark superfields with
fermion components $\psi, \bar\psi$. $m$ is the meson superfield,
it is related by duality to the composite field $Q\bar Q$ of the
eletric theory. The fermionic components of $m$ will be
denoted by $\chi$. The magnetic theory also has a tree level
superpotential $W=m q \bar q$.

Two discrete subgroups of the global and gauge symmetries
given above will be important for the projection which we
perform in the next subsection. Both correspond to continuous
$U(1)_R$ symmetries in the classical limit but are broken
to $Z_{3N}$ and $Z_{3(F-N)}$ by instantons in the quantum theory. 
The charges of the various fields under the two symmetries are
given in the following table.

\beq
\begin{array}{l||rrrrr||rrrrrrr}
&\lambda&Q&\bar Q&\Psi&\bar\Psi&\twi\lambda&q&\bar q&\psi&\bar\psi&m&\chi
\\[.1in]\hline &&&&&&&&&&&&\\[-.1in]
Z_{3N}&1&1&1&0&0&1&0&0&-1&-1&2&1 \\ 
Z_{3(F-N)}&1&0&0&-1&-1&1&1&1&0&0&0&-1 \\
\end{array}
\eqn{discreteqcd}
\eeq
\medskip
Note that the role of the two symmetries is exchanged under duality.

There are a number of consistency checks which have been performed
on this duality. One check is that there exists an operator
map between gauge invariant chiral operators which is consistent with
all the global symmetries.
\begin{eqnarray*}
Q\bar Q &\leftrightarrow & m \\
Q^N     &\leftrightarrow & q^{F-N} \\
\bar Q^N  &\leftrightarrow & \bar q^{F-N} 
\end{eqnarray*}
The global anomalies of both theories match, both theories
have the same moduli space of vacua, and finally the duality
is consistent under deformations of the theory.

In the following subsections we perform orbifold projections
on both electric and magnetic theories. At leading order in large
N we expect the orbifolded theories' correlators to be given by
their supersymmetric parent theories' modulo the
rescaling of the coupling constants. Since the parent theories
are dual in the IR, we conclude that the daughters are also
dual.

\subsection{\it Projecting the electric theory}

The projection is carried out by identifying a $Z_3$ subgroup
of the global symmetries and then projecting out all the fields
which are not $Z_3$ invariant. As explained in the main body
of the paper the $Z_3$ symmetry is taken to act as a multiple
of the regular representation on the $SU(3N)$ and $SU(3F)$ indices.
In addition it may also be mixed with any of the finite global
symmetries. For example it could be acting trivially on all global
charges, it could be embedded into Baryon number $B$ or the $R$
symmetry, or into one of the two discrete symmetries.
If the projection
involves a subgroup of the $R$ symmetry, fermions and
bosons are projected differently and supersymmetry
is broken.

I will first describe the action on $SU(3N)$ and $SU(3F)$
and then discuss the different possibilities for the finite
global charges.
$SU(3N)$ and $SU(3F)$ indices transform in multiples
of the regular representation. The regular representation
of $Z_3$ is given by the matrices

\beq
\Biggl\{\left(\matrix{
1&0&0 \cr 0&1&0 \cr 0&0&1 \cr}
\right)\ , 
\left(\matrix{
1&0&0 \cr 0&\omega&0 \cr 0&0&\omega^2 \cr}
\right)\ , 
\left(\matrix{
1&0&0 \cr 0&\omega^2&0 \cr 0&0&\omega \cr}
\right)\Biggr\}\ ,\quad \omega=e^{2\pi i/3}\  . 
\eeq

It suffices to check invariance under the second group element
above because it generates the group. We take an $SU(3N)$
index to transforms by multiplication with a matrix in an
N-fold copy of the regular representation. For the second group
element this $3N\times3N$ matrix is 

\beq
\Omega_{3N}=\left(\matrix{\one_N&|&&|&\cr --&--&--&--&-- \cr
               &|&\omega\one_N&|&\cr --&--&--&--&-- \cr
               &|&&|&\omega^2\one_N\cr}
\right)
\eeq

where $\one_N$ is the $N\times N$ unit matrix.
The gluons in the adjoint representation of $SU(3N)$ then
transform as

\beq
A_\mu \longrightarrow \Omega_{3N} A_\mu \Omega_{3N}^\dagger \ ,
\eeq
It is clear that only the block-diagonal components of
$A_\mu$ are invariant. Thus the projection breaks the
$SU(3N)$ to its $SU(N)\times SU(N) \times SU(N)$
subgroup.

The projection of the matter fields depends on the chosen
embedding of the $Z_3$ into the global symmetries.
As a warm-up we first discuss the case where the only
$Z_3$ transformation comes from the gauge and $SU(3F)$
indices:

In this case supersymmetry will be unbroken. Thus gauginos
are projected identically to gauge bosons.
The quarks and squarks both transform as $3N,\bar{3F}$
under $SU(3N)\times SU(3F)$. Transforming the fundamental
color index with $\Omega_{3N}$ and the antifundamental flavor
index with the $3F \times 3F$ dimensional
$\Omega_{3F}^\dagger$ we find that the invariant fields
are $N\times F$ blocks on the diagonal of $Q$
 
\beq
\eqn{elecquarks}
\pmatrix{Q_1&|&&|&\cr --&--&--&--&-- \cr
               &|&Q_2&|&\cr --&--&--&--&-- \cr
               &|&&|&Q_3\cr} \ .
\eeq

Similarly, only the block-diagonal components of $\bar Q$ survive
the projection. The interactions of the daughter theory
are obtained from the parent theory by simply dropping any
interaction terms involving non-$Z_3$ invariant fields.

The result is three completely decoupled identical
copies of supersymmetric QCD with
gauge groups $SU(N)$ and flavor groups $SU(F)\times SU(F)$.
It is clear that we will not gain any new insight into the
dynamics of supersymmetric QCD  by studying this projection
since it just relates supersymmetric QCD to itself modulo
a re-scaling of the number of colors and flavors.

We will show in the next subsection that the dual theory
is also simply projected down to three independent copies
of $SU(F-N)$ gauge theory. Thus we see that the projection
commutes with duality in this case.

Let us now discuss a non-trivial projection which breaks
supersymmetry. We embed the $Z_3$ in the discrete $Z_{3N}$
$R$-symmetry. In addition to the transformation from the
regular representation each field is multiplied by
$\omega$ to the power of its $Z_{3N}$ charge (see table).
For example, the gauginos transform with an additional factor
of $\omega$
compared to the gauge bosons $\lambda\rightarrow \omega \ 
\Omega_{3N} \lambda \Omega_{3N}^\dagger$. The invariant gauginos in
$N\times N$ block-notation are
\beq
\pmatrix{&\lambda_1& \cr &&\lambda_2 \cr \lambda_3&&\cr} \ .
\eeq

Squarks and quarks transform as $Q\rightarrow \omega\Omega_{3N}
Q\Omega_{3F}^\dagger$ and $\Psi\rightarrow\Omega_{3N}^\dagger
\Psi\Omega_{3F}$.
All matter fields which survive the projection are given
in the following table.

\beq
\begin{array}{c|ccc|cccccc|rr}
&\smsu N&\smsu N&\smsu N&\smsu F&\smsu F&\smsu F&\smsu F&\smsu F&\smsu F&
B&R \\[.1in] \hline 
\lambda_1 & \Ybox &\Ybbox &   &&& &&& &0&1  \\
\lambda_2 &  &\Ybox & \Ybbox  &&& &&& &0&1 \\
\lambda_3 & \Ybbox & & \Ybox  &&& &&& &0&1 \\[.1in]

Q_1 &\Ybox & & &  & \Ybbox  &  &&& &{1\over N}&{F-N\over F} \\
Q_2 & &\Ybox & &  &  & \Ybbox  &&& &{1\over N}&{F-N\over F} \\
Q_3 & & &\Ybox & \Ybbox  &  &  &&& &{1\over N}&{F-N\over F} \\[.1in]

\bar Q_1& \Ybbox & & & &&& & & \Ybox &-{1\over N}&{F-N\over F} \\
\bar Q_2& & \Ybbox & & &&& \Ybox & & &-{1\over N}&{F-N\over F} \\
\bar Q_3& & & \Ybbox & &&& & \Ybox & &-{1\over N}&{F-N\over F} \\[.1in]

\Psi_1& \Ybox & & & \Ybbox & & &&& &{1\over N}&{-N\over F} \\
\Psi_2& & \Ybox & & & \Ybbox & &&& &{1\over N}&{-N\over F} \\
\Psi_3& & & \Ybox & & & \Ybbox &&& &{1\over N}&{-N\over F} \\[.1in]

\bar\Psi_1 & \Ybbox & & & &&& \Ybox & & &\!-{1\over N}&{-N\over F} \\
\bar\Psi_2 & & \Ybbox & & &&& & \Ybox & &\!-{1\over N}&{-N\over F} \\
\bar\Psi_3 & & & \Ybbox & &&& & & \Ybox &\!-{1\over N}&{-N\over F} \\

\end{array}
\eqn{electric}
\nonumber
\eeq
\medskip

The Lagrangian of the orbifolded theory contains the usual
kinetic terms for all the fields with gauge covariant derivatives.
In addition there are Yukawa and scalar couplings of the form
\beq
\eqn{gauginomatter}
\CL \sim \sum_i \left[Q_i^\dagger \lambda_i \Psi_{i+1} +
\bar Q_i^\dagger \lambda_{i-1} \bar \Psi_{i-1}
+ \sum_a \left|Q_i^\dagger t^a Q_i
-\bar Q_i^\dagger t^a \bar Q_i\right|^2\right] \ ,
\eeq
where $t^a$ are generators of the $SU(N)$ algebra, gauge and
flavor indices are traced over. In canonical normalization for all
fields the scalar couplings are equal to $g^2$ and gauge and Yukawa
couplings are $g$.

\subsection{\it Projecting the magnetic theory and duality}

To orbifold the magnetic theory we again assign gauge indices to
transform in the regular representation\footnote{
Note that since the regular representation is real, it does
not matter whether we assign fundamental or antifundamental
indices to transform in the regular representation.}.
In order to decide how to treat
global indices we need to copy exactly what we did in the
electric theory. The magnetic theory  has the same global symmetries
as the electric theory, that uniquely fixes the projection.

Performing the projection as described in the previous subsection
we find --not very surprisingly -- that the dual gauge group is projected as
$$SU(3\twi N)\rightarrow SU(\twi N)\times SU(\twi N)\times SU(\twi N)\ .$$

The projection on gluinos and quarks depends on our choice of embedding
for $U(1)_R$. Recall the warm-up example where we embedded the
${\bf Z}_3$ trivially into $U(1)_R$. Doing the same in the dual
also preserves supersymmetry; thus we have gluinos
in the same representations as gluons, and dual quarks are projected
similarly to electric quarks \eq{elecquarks}. The new
ingredient is the meson field $m$: it also simply gets projected
to three blocks on the diagonal. 
Thus we find three decoupled copies of $SU(\twi N)$ theories
with dual quarks and meson fields for each. This is obviously
the dual of three copies of supersymmetric QCD with gauge
group $SU(N)$. Thus we find no surprises, the orbifold projection
applied to the electric-magnetic $SU(3N)\leftrightarrow
SU(3(F-N))$ dual pair yielded three times an $SU(N)
\leftrightarrow SU(F-N)$ pair.

Much more interesting is the non-trivial projection discussed
in the electric theory. Again the dual gauge group gets
projected to $SU(\twi N)\times SU(\twi N)\times SU(\twi N)$.
The projection of the matter fields is also straightforward
using the charge assignments given in \eq{magneticqcd} and \eq{discreteqcd}.

\beq
\begin{array}{c|ccc|cccccc|rr}
&\smsu {\twi N}&\smsu {\twi N}&\smsu {\twi N}&
\smsu F&\smsu F&\smsu F&\smsu F&\smsu F&\smsu F&
B&R \\[.1in] \hline 
\twi\lambda_1 & \Ybox &\Ybbox &   &&& &&& &0&1  \\
\twi\lambda_2 &  &\Ybox & \Ybbox  &&& &&& &0&1 \\
\twi\lambda_3 & \Ybbox & & \Ybox  &&& &&& &0&1 \\[.1in]

q_1 & \Ybbox & &   & \Ybox &   &   &&& &{1\over\twi N}&{N\over F} \\
q_2 &  &\Ybbox &   &   & \Ybox  &  &&& &{1\over\twi N}&{N\over F} \\
q_3 &  & & \Ybbox  &   &  & \Ybox  &&& &{1\over\twi N}&{N\over F} \\[.1in]

\bar q_1& \Ybox & & & &&&\Ybbox & & &-{1\over\twi N}&{N\over F} \\
\bar q_2& & \Ybox & & &&& &\Ybbox & &-{1\over\twi N}&{N\over F} \\
\bar q_3& & & \Ybox & &&& & &\Ybbox &-{1\over\twi N}&{N\over F} \\[.1in]

\psi_1& \Ybbox & & & & \Ybox & &&& &{1\over\twi N}&{N-F\over F} \\
\psi_2& & \Ybbox & & & & \Ybox &&& &{1\over\twi N}&{N-F\over F} \\
\psi_3& & & \Ybbox & \Ybox & & &&& &{1\over\twi N}&{N-F\over F} \\[.1in]

\bar\psi_1 &\Ybox & & & &&& & &\Ybbox &-{1\over\twi N}&{N-F\over F} \\
\bar\psi_2 & &\Ybox & & &&&\Ybbox & & &-{1\over\twi N}&{N-F\over F} \\
\bar\psi_3 & & &\Ybox & &&& &\Ybbox & &-{1\over\twi N}&{N-F\over F} \\[.1in]

m_1 &&&& & \Ybbox & & & & \Ybox & 0&2{F-N\over F} \\
m_2 &&&& & & \Ybbox & \Ybox & & & 0&2{F-N\over F} \\
m_3 &&&& \Ybbox & & & & \Ybox & & 0&2{F-N\over F} \\[.1in]

\chi_1 &&&& & & \Ybbox & & \Ybox & &0&1-2{N\over F} \\
\chi_2 &&&& \Ybbox & & & & & \Ybox &0&1-2{N\over F} \\
\chi_3 &&&& & \Ybbox & & \Ybox & & &0&1-2{N\over F} \\

\end{array}
\eqn{magnetic}
\nonumber
\eeq
\medskip

This theory has a slightly more complicated Lagrangian than the
electric theory due to the remains of the superpotential term.
We find

\begin{eqnarray}
\eqn{dualyuka}
\CL &\sim& \sum_i\left[ q_i^\dagger \twi\lambda_{i-1} \psi_{i-1}
+ \bar q_i^\dagger \twi\lambda_i \bar\psi_{i+1}
+ \sum_a \left|q_i^\dagger t^a q_i -\bar q_i^\dagger t^a \bar q_i \right|^2
\right. \\
&&\left.
+ \psi_i m_i\bar\psi_i + q_i\chi_{i+2}\bar\psi_i + \psi_i\chi_{i-1}\bar q_i
+ \left|q_i \bar q_i\right|^2 + \left|q_i m_{i-1}\right|^2
+ \left|\bar q_i m_{i+1} \right|^2 \right]\ . \nonumber
\end{eqnarray}

Note that this dual is weakly coupled in the infrared for
$F \le {3\over2}N$. Completely analogous to the parent
theory duality this duality also exchanges strong and
weak coupling.
 
We postpone performing a full set of consistency checks on our
proposed duality to future work. Here we just give the map of
gauge invariants containing only scalars
\begin{eqnarray*}
Q_i\bar Q_i &\leftrightarrow & m_i \\
Q_i^N     &\leftrightarrow & q_{i+1}^{F-N} \\
\bar Q_i^N  &\leftrightarrow & \bar q_{i-1}^{F-N} \ .
\end{eqnarray*}
This map is necessary to perform a comparison of the flat directions in
the two theories.

It is straightforward to compare the anomalies
in both eletric and magnetic theories. All the anomalies for
the global symmetries given in the tables agree. This also follows
from our derivation of the dual pair as follows: the anomalies
mentioned above are all calculated with a planar triangle diagram
with global currents at the vertices. According to the general
arguments presented in the previous section, these diagrams
are related to the corresponding anomaly diagrams in the parent
theories (supersymmetric QCD and it's dual) via rescaling by
a factor of $\Gamma=3$. Since the anomalies matched in the
electric and magnetic descriptions of supersymmetric QCD,
and since the global group of the daughter theories is a
subgroup of the global symmetries of the parents, the anomalies
must match here as well. Note that this derivation makes it clear
that the anomalies also match for finite N which might be of
significance for attempts to continue the proposed duality
to finite N.

\section{Acknowledgements}
I am grateful to Josh Erlich and Asad Naqvi for collaboration during
the initial stage of this work and for helpful discussions. 
I am also imdebted to Andrew Cohen and Raman Sundrum for sharing their
wisdom with me on many occasions. 
This research is supported by DOE grant \#DE-FG02-91ER40676.


\begin{thebibliography}{99}

\bibitem{seiberg}
N. Seiberg, `` Exact Results on the Space of Vacua of four-dimensional
SUSY Gauge Theories,'' \PRD{49}{6857}{1994}, hep-th/9402044;
``Electric-Magnetic Duality in Supersymmetric non-Abelian Gauge Theories,''
\NPB{435}{129}{1995}, hep-th/9411149.

\bibitem{intrseib}
{\it for reviews see e.g.}
K. Intriligator and N. Seiberg, ``Lectures on Supersymmetric Gauge
Theories and Electric-Magnetic Duality,''
{\it Nucl. Phys. Proc. Suppl.} {\bf 45BC}, 1 (1996), hep-th/9509066;
M.E. Peskin, ``Duality in Supersymmetric Yang-Mills Theory,'' hep-th/9702094;
M.A. Shifman, ``Nonperturbative Dynamics in Supersymmetric Gauge Theories,''
{\it Prog. Part. Nucl. Phys.} {\bf 39} 1 (1997), hep-th/9704114.

\bibitem{john}
J. Terning, ``t Hooft Anomaly Matching for QCD,'' \PRL{80}{2517}{1998},
hep-th/9706074.

\bibitem{nbranes}
A. Brandhuber, J. Sonnenschein, S. Theisen, and S. Yankielowicz,
``Brane Configurations and 4-D Field Theory Dualities,''
\NPB{502}{125}{1997}, hep-th/9704044;
N. Evans and M. Schwetz, ``The Field Theory of Nonsupersymmetric
Brane Configurations,'' hep-th/9708122.

\bibitem{malda}
J. Maldacena, ``The large N Limit of Superconformal Field Theories and
Supergravity,'' hep-th/9711200.

\bibitem{KS}
S. Kachru and E. Silverstein, ``4-D conformal Theories and Strings on
Orbifolds,'' hep-th/9802183.

\bibitem{harvard}
A. Lawrence, N. Nekrasov, and C. Vafa, ``On Conformal Field Theories in
four-Dimensions,'' hep-th/9803015.
M. Bershadsky, Z. Kakushadze,and C. Vafa, ``String Expansion as large
N Expansion of Gauge Theories,'' hep-th/9803076.
Z. Kakushadze, ``Gauge Theories from Orientifolds and large N Limit,''
hep-th/9803214.

\bibitem{BJ}
M. Bershadsky and A. Johansen, ``Large N Limit of Orbifold Field Theories,''
hep-th/9803249.

\bibitem{etaed}
E. Witten, ``Instantons, the Quark Model, and the 1/N Expansion,''
\NPB{149}{285}{1979}.

\bibitem{ds}
M.R. Douglas and S.H. Shenker, ``Dynamics of SU(N) Supersymmetric Gauge
Theory,'' \NPB{447}{271}{1995}, hep-th/9503163.

\bibitem{shifvain}
M.A. Shifman and A.I. Vainshtein, ``On Gluino Condensation in Supersymmetric
Gauge Theories, SU(N) and O(N) Groups,'' \NPB{296}{445}{1988}.

\bibitem{josh}
K. Intriligator and N. Seiberg, ``Phases of N=1 Supersymmetric Gauge
Theories in Four-Dimensions,'' \NPB{431}{551}{1995}, hep-th/9408155.
C. Csaki, J. Erlich, D. Freedman, and W. Skiba, ``N=1 Supersymmetric
Product Group Theories in the Coulomb Phase,'' \PRD{56}{5209}{1997},
hep-th/9704067.

\bibitem{erich}
J. Lykken, E. Poppitz, and S.P. Trivedi, ``Chiral Gauge Theories
from D-Branes,'' \PLB{416}{286}{1998}, hep-th/9708134; ``M(ore)
on Chiral Gauge Theories from D-Branes,'' hep-th/9712193.

\bibitem{thoo}
G. 't Hooft, ``A Planar Diagram Theory for Strong Interactions,''
\NPB{72}{461}{1974}.

\bibitem{nima}
N. Arkani-Hamed and H. Murayama, ``Holomorphy, Rescaling Anomalies and exact
Beta Functions in Supersymmetric Gauge Theories,'' hep-th/9707133.

\bibitem{wittenmqcd}
E. Witten, ``Branes and the Dynamics of QCD,'' \NPB{507}{658}{1997},
hep-th/9706109.

\bibitem{shif}
M.A. Shifman and M.B. Voloshin, ``Degenerate Domain Wall Solutions in
Supersymmetric Theories,'' \PRD{57}{2590}{1998}, hep-th/9709137.

\bibitem{Sidney}
S. Coleman, ``1/N,'' in {\it Aspects of Symmetry}, Cambridge University
Press, Cambridge, 1985.

\bibitem{Howard}
H. Georgi, ``A Tool Kit for Builders of composite Models,''
\NPB{266}{274}{1986}.

\bibitem{nick}
N. Evans and M. Schmaltz, ``A diagrammatic Analysis of Duality in
Supersymmetric Gauge Theories,'' \PRD{55}{3776}{1997}, hep-th/9609183.

\bibitem{quivers}
M.R. Douglas and G. Moore, ``D-branes, Quivers, and ALE Instantons,''
hep-th/9603167.



\end{thebibliography}
\end{document}